\documentclass[showpacs,twocolumn,prl,floatfix,superscriptaddress]{revtex4}
\usepackage{graphicx}
\usepackage{amsmath,amsfonts}
\usepackage{amssymb}

\begin{document}
\title{Stimulated Raman spin-coherence and spin-flip induced hole burning in charged GaAs quantum dots}
\author{Jun Cheng}
\altaffiliation[Present Address: ]{Department of Biomedical
Engineering, University of Michigan, Ann Arbor, MI 48109}
\affiliation{The H. M. Randall Laboratory of Physics,
            The University of Michigan,
            Ann Arbor, MI 48109}
\author{Wang Yao}
\altaffiliation[Present Address: ]{Department of Physics, The
University of Texas, Austin, TX 78712}
\affiliation{Department of
Physics,
        The University of California-San Diego,
        La Jolla, CA 92093}
\author{Xiaodong Xu}
\author{D. G. Steel}
\email{dst@umich.edu}
\affiliation{The H. M. Randall Laboratory of Physics,
            The University of Michigan,
            Ann Arbor, MI 48109}
\author{A. S. Bracker}\author{D. Gammon}
\affiliation{The Naval Research Laboratory,
            Washington D. C. 20375}
\author{L. J. Sham}
\affiliation{Department of Physics,
        The University of California-San Diego,
        La Jolla, CA 92093}


\begin{abstract}
High-resolution spectral hole burning (SHB) in coherent
non-degenerate differential transmission spectroscopy discloses
spin-trion dynamics in an ensemble of negatively charged quantum
dots. In the Voigt geometry, stimulated Raman spin coherence gives
rise to Stokes and anti-Stokes sidebands on top of the trion
spectral hole. The prominent feature of an extremely narrow spike
at zero detuning arises from spin population pulsation dynamics.
These SHB features confirm coherent electron spin dynamics in
charged dots and the linewidths reveal spin spectral diffusion
processes.
\end{abstract}

\pacs{71.35.Pq, 42.65.-k, 78.67.Hc}

\maketitle

Single electron spin localized in semiconductor quantum dots (QDs)
has attracted a great deal of interest due to its potential use in
quantum applications \cite{Awschalom_book}. Experimental and
theoretical efforts have been focused on controllable coherent
spin dynamics and possible decoherence mechanisms
\cite{Dutt2005,Petta_Science2005,Koppens_Rabi,Fujisawa_2002,HansonPRL03,Elzerman_Nature2004,Kroutvar_Nature2004,Khaetskii_2001,Merkulov_2002,Cheng_SSC2006,
Loss_SpinT2_phonon,Yao_PRB2006,Espin_HF_dipole_2_DaSSarma}. In
this paper, we report spectral hole burning (SHB) in coherent
differential transmission (DT) spectroscopy induced by spin-trion
dynamics in an ensemble of negatively charged QDs. Spin coherence
induced SHB by stimulated Raman excitation and spin relaxation
induced SHB due to the population pulsation dynamics are observed
in the QD system. Features of SHB not only disclose important spin
dynamics observed from transient spectroscopy~\cite{Dutt2005} and
phase modulation techniques~\cite{Cheng_SSC2006}, but also provide
information on spin spectral diffusion~(SD) processes
\cite{Espin_HF_dipole_2_DaSSarma} which are not easily revealed by
previous methods.

The interface fluctuation GaAs/Al$_{0.3}$Ga$_{0.7}$As QDs are
molecular beam epitaxy grown with growth interrupts, and
modulation Si doping in the barrier incorporates excess
electrons~\cite{Gammon_PRB2002}. The pump $E_{1}$($\omega_{1}$)
and probe $E_{2}$($\omega_{2}$) optical fields are derived from
two frequency-stabilized and independently tunable CW lasers with
a mutual coherence bandwidth of 20 neV, which is crucial for this
experiment as discussed below.
\begin{figure}[tb] \centering
\includegraphics[width=3.4in]{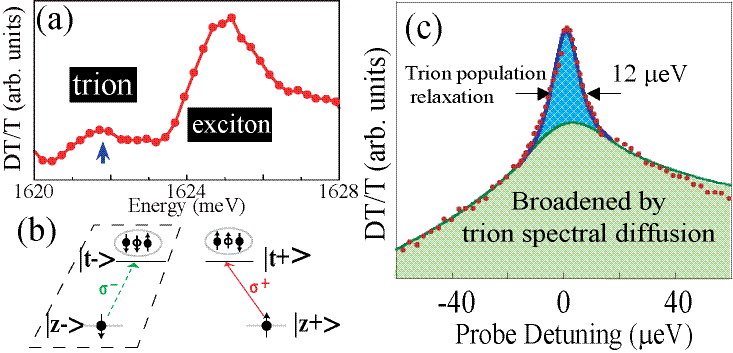}
\caption{(a). Coherent degenerate DT spectrum of both trion and
exciton ensemble resonances. The arrow shows the fixed pumping
position for (c). (b). Energy level diagram of charged QDs in
absence of magnetic field, where solid (empty) circles indicate
electrons (holes) and arrows show the corresponding spin
orientations. The single electron ground state has either spin
down ($|z-\rangle$) or spin up ($|z+\rangle$). The trion state
consists of two-electron singlet state and one heavy hole with
spin down ($|t-\rangle$) or spin up ($|t+\rangle$). $\sigma^-$
($\sigma^+$) light couples $|z-\rangle$ ($|z+\rangle$) to
$|t-\rangle$ ($|t+\rangle$). (c) A clear view of the trion
spectral hole burning profile, shown as a double Lorentzian, where
the probe is detuned relative to pump position (1621.9 meV).
} \label{fig_noB-data}
\end{figure}
The sample is kept inside a superconducting magnetic liquid helium
flow cryostat and the temperature is maintained at 4.5 K. The DT
signal is homodyne detected with the probe field by a photodiode
and extracted by a lock-in amplifier.

A nonlinear degenerate DT spectrum with $\omega_1=\omega_2$
[Fig.~\ref{fig_noB-data}(a)] shows the trion and exciton ensemble
resonances. Their assignments are confirmed by both
photoluminescence and transient quantum beats studies (data not
shown). The trion binding energy (i.e., the separation between
exciton and trion resonances) is measured to be 2.9 meV, in
agreement with earlier reports of
photoluminescence~\cite{Gammon_PRB2002} and transient
spectroscopy~\cite{Dutt2005}. The ensemble trion inhomogeneous
broadening width ($\sim 2.5$ meV) can be estimated from the broad
Gaussian profile of the trion resonance.

\begin{figure}[tb] \centering
\includegraphics[width=3in]{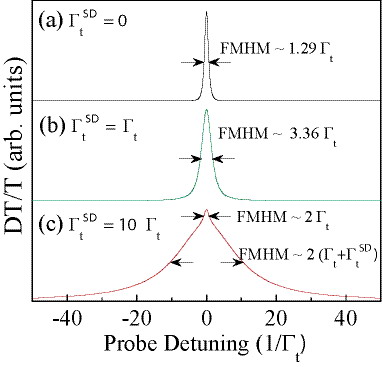} \caption{The calculated SHB lineshape of 2-level system with different
spectral diffusion rates, where $\Lambda_t=1000\Gamma_t$ and
$\gamma_t=\Gamma_t/2$. (a). $\Gamma_t^{SD}=0$. (b).
$\Gamma_t^{SD}=\Gamma_t$. (c). $\Gamma_t^{SD}=10\Gamma_t$.}
\label{fig_noB-theory}
\end{figure}

To study SHB with non-degenerate DT spectroscopy, the pump beam is
fixed near the trion ensemble peak and the probe beam is detuned
($\Delta\equiv\omega_2-\omega_1$). A narrow spectral structure
appears, exhibiting a double-Lorentzian-like shape, i.e., a
narrower Lorentzian peak on top of a broader one
[Fig.~\ref{fig_noB-data}(c)]. As established below, the narrower
Lorentzian peak is due to the trion population relaxation dynamics
and its linewidth ($\sim 12~\mu$eV) gives the trion population
relaxation rate. The broader Lorentzian peak ($\sim 50~\mu$eV)
underneath this resonance is due to trion population relaxation
and coherence decay broadened by the trion SD process. This
complex lineshape and its unfolded SHB features with a magnetic
field in the Voigt geometry are the focus of our study.

The dynamics of the QD ensemble in the presence of the radiation
field and various interactions with the environment is described
by the modified optical Bloch equation~(OBE)~\cite{Berman_1985}:
\begin{equation}
 \label{master_equation}
   i \hbar  \dot{\rho}(\epsilon) = [H,\rho(\epsilon)] + \frac{\partial \rho (\epsilon)}{\partial t}
    \bigg | _{\text{relax}} + \frac{\partial \rho (\epsilon)}{\partial t}
    \bigg | _{\text{SD}}
\end{equation}
where $\rho(\epsilon)$ is the density matrix of each ensemble
member characterized by its resonance energy $\epsilon$. $H$ is
the total Hamiltonian including the interaction with the coherent
optical fields. The second term on the RHS is a generalized
relaxation term that describes population decay and pure
dephasing. The last term is due to various SD
processes~\cite{Meystre_Sargent_book}. Eq.~(\ref{master_equation})
without the SD term reduces to the standard OBE.

Considering the optical selection rules in the absence of a
magnetic field, the transition from spin ground state $|z-\rangle$
($|z+\rangle$) to trion state $|t+\rangle$ ($|t-\rangle$) is
forbidden as a dark transition~\cite{Gammon_PRB2002}. Therefore,
with co-circularly polarized pump and probe fields, the charged QD
energy levels reduce to a 2-level system, as shown in the enclosed
box in Fig.~\ref{fig_noB-data}(b). The relevant SD process here is
the interdot transfer of trion population
\cite{Wang_Hailin_PRL1990}
\begin{eqnarray}
 \label{eq:2-level}
\frac{\partial \rho_{\bar{t},\bar{t}} (\epsilon)}{\partial t}
    \bigg | _{\text{SD}} &=&-\Gamma_t^{SD}(\epsilon)\rho_{\bar{t},\bar{t}}(\epsilon)
    +\int W_t(\epsilon,\epsilon')
    \rho_{\bar{t},\bar{t}}(\epsilon')d\epsilon'
    , \nonumber \\
\frac{\partial \rho_{\bar{t},\bar{z}} (\epsilon)}{\partial t}
    \bigg | _{\text{SD}} &=&
    -\Gamma_t^{SD} (\epsilon) \rho_{\bar{t},\bar{z}}(\epsilon).
\end{eqnarray}
where $W_t(\epsilon,\epsilon')$ is the spectral redistribution
kernel representing the rate for trion population to migrate from
a dot with resonant energy $\epsilon'$ to dots with energy
$\epsilon$. While this interdot transfer conserves the ensemble
trion population, the dipole coherence is lost.
$\Gamma_t^{SD}(\epsilon)=\int W_t(\epsilon',\epsilon) d\epsilon'$
is the overall SD rate. For an optically thin sample, the DT
spectrum is given by the imaginary part of the third order induced
optical polarization integrated over the inhomogeneous
distribution~\cite{Wang_Hailin_PRA1991,NicoPRL98},
\begin{widetext}
\begin{eqnarray}
P_{\textrm{NL}}^{(3)} &\simeq& - \frac{iN|\mu|^4|E_1|^2 E_2^*}{2\hbar^3\Lambda_t} \Bigg \{
\frac{1}{\Gamma_t+\Gamma_t^{SD}+i\Delta} \left[ \frac{1}{\Lambda_t}
+ \frac{\sqrt{\pi}}{2(\gamma_t+\Gamma_t^{SD})+i\Delta}\right] +
\frac{\pi}{\Gamma_t\Lambda_t}\frac{\Gamma_t^{SD}}{\Gamma_t+\Gamma_t^{SD}}
\nonumber \\
&& +\frac{\sqrt{\pi}}{\Gamma_t+\Gamma_t^{SD}}
\frac{1}{2(\gamma_t+\Gamma_t^{SD})+i\Delta}
+\frac{\pi}{\Lambda_t(\Gamma_t+i\Delta)}
\frac{\Gamma_t^{SD}}{\Gamma_t+\Gamma_t^{SD}+i\Delta} \Bigg \}
\label{eq:SHB-sd2l-finalP}
\end{eqnarray}
\end{widetext}
where $\Gamma_t$ ($\gamma_t$) is trion population (coherence)
decay rate, $N$ is the total number of excited charged QDs,
$\Lambda_t$ is the ensemble inhomogeneous broadening width of the
trion resonance and $\mu$ the optical dipole of a single dot. We
have assumed a redistribution kernel $W_t(\epsilon,\epsilon') =
\Gamma_t^{SD}
\text{exp}(-(\epsilon-\bar{\epsilon})^2/(\Lambda_t)^2)/(\sqrt{\pi}
\Lambda_t)$ where $\bar{\epsilon}$ is the ensemble averaged
resonance energy \cite{SD_kernel}. Eq.~(\ref{eq:SHB-sd2l-finalP})
holds when the approximation of plasma dispersion function is
taken~\cite{Fried_book} because $\Gamma_t,\gamma_t,\Gamma_t^{SD},
\Delta \ll \Lambda_t$ and the pump frequency is fixed near the
center of the ensemble trion spectrum.

The SD process significantly changes the trion SHB lineshape and
linewidth, which can be discussed in three regimes, as
schematically shown in Fig.~\ref{fig_noB-theory}.
$\Gamma_t=2\gamma_t$ is assumed only for the theoretical
calculation in Fig.~\ref{fig_noB-theory} to simplify discussions,
since pure dephasing has been found negligible for trion states at
$4.5$ K~\cite{Dutt_SSC2006}. When $\Gamma_t^{SD} \ll \Gamma_t$,
the trion SHB in the DT spectrum reduces to the standard
Lorentzian squared with width $\sim \Gamma_t$. Secondly, when
$\Gamma_t^{SD} \simeq \Gamma_t$, the trion SHB lineshape is
$(24\Gamma_t^2+\Delta^2)
[(\Gamma_t+\Gamma_t^{SD})^2+\Delta^2]^{-1}[(\Gamma_t+2\Gamma_t^{SD})^2+\Delta^2]^{-1}$
where the linewidth is considerably broadened by the SD process.
Finally, when $\Gamma_t^{SD}\gg\Gamma_t$, the lineshape changes to
a double Lorentzian-like profile with larger FWHM of
$2(\Gamma_t+\Gamma_t^{SD})$ and smaller FWHM of $2\Gamma_t$. The
physical origin of the narrower Lorentzian is due to the
population pulsation effect \cite{Duncan_PRL1985}. This explains
the observed lineshape shown in Fig.~\ref{fig_noB-data}(c), which
is consistent with the physical model
(Eq.~(\ref{eq:SHB-sd2l-finalP})) assuming $\hbar\Gamma_t \simeq 7
\mu$eV, $\hbar\gamma_t \simeq 8 \mu$eV, $\hbar\Gamma_t^{SD} \simeq
46\mu$eV, and $\hbar\Lambda_t \simeq 1000 \mu$eV plus a slow
linear slope. The values of $\Gamma_t$ and $\gamma_t$ agree with
earlier reports~\cite{Dutt2005,Dutt_SSC2006} and the value of
$\Lambda_t$ is in agreement with the degenerate DT trion ensemble
resonance in Fig.~\ref{fig_noB-data}(a). The relatively big
$\Gamma_t^{SD}$ implies trion SD process plays an important role.

\begin{figure}[tb] \centering
\includegraphics[width=3.4in]{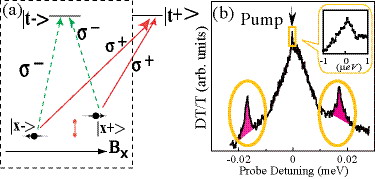}
 \caption{(a). Energy level scheme of charged dot in Voigt geometry ($B_x\neq0$).
The new ground states $| x \pm \rangle$ denote electron spin
orientation along the $x$-axis separated by Zeeman splitting. The
new selection rules are labelled by solid (dashed) lines with
$\sigma^{+}(\sigma^{-})$ light. (b). The newly-emerged center
sharp spike and two symmetric sidebands at $B_x$ = 2.2 T, where
the fixed pump is indicated by the arrow. The two Stokes and
anti-Stokes spin-coherence induced sidebands are highlighted. The
inset shows the zoomed in sharp central spike at $\Delta \sim 0$.
A similar figure was first presented in Ref.~\cite{Cheng_SSC2006}
but with a different objective.} \label{fig_voigt-level}
\end{figure}

With a magnetic field applied perpendicular to QD growth direction
(i.e., Voigt geometry), the spin eigenstates $|x \pm \rangle$
along the field direction are split in energy due to the non-zero
electron in-plane g-factor ($g_x^e$) while the trion states are
unaffected as the heavy-hole in-plane g-factor is
negligible~\cite{DanPRB2002}. The new eigenstates and optical
selection rules are shown in
Fig.~\ref{fig_voigt-level}(a)~\cite{Dutt2005}. With co-circularly
polarized pump and probe fields, the relevant states form a
three-level $\Lambda$-system enclosed in the dashed box. Utilizing
a similar $\Lambda$-system, optical pumping leading to spin
cooling has been reported recently in different structural
QDs~\cite{Atature_Science2006, Xu_spincooling}. However, no
similar optical pumping effect was observed in our system, where
the strong spectral diffusion on two spin ground states discussed
below plays an important role.

\begin{figure}[tb] \centering
\includegraphics[width=3.4in]{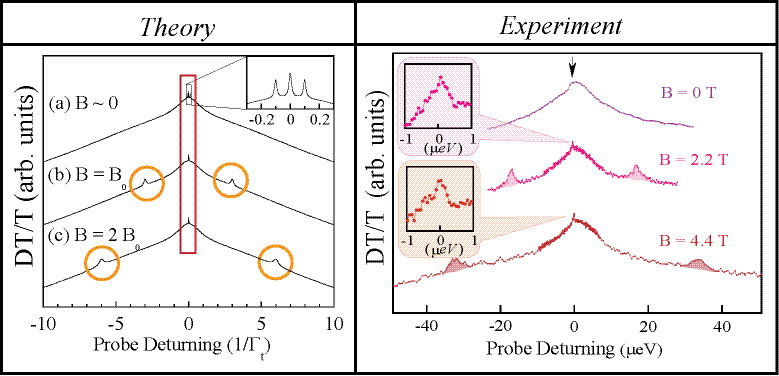} \caption{Comparison
of theoretical calculations and experimental results in Voigt
geometry. {\it Theory}: The DT spectrum of 3-level $\Lambda$-system
at different magnetic field, where $\Lambda_t = 1000 \Gamma_t$,
$\gamma_t\simeq\Gamma_t/2$, and $\Gamma_s^{SD}=0.01\Gamma_t$. (a).
$B\sim 0$ so that spin inhomogeneous broadening is due to nuclear
field $\Lambda_s \sim \Lambda_s^n$. The central spike and the
sidebands merge in the limit of vanishing $B$. (b). $B=B_0$ where
$\mu_B \bar{g}^x_e B_0=3 \hbar \Gamma_t$. Spin inhomogeneous
broadening is dominated by the inhomogeneity of the g-factor:
$\Lambda_s \simeq \mu_B \Delta g^x_e B /\hbar$. (c). $B=2B_0$. {\it
Experiment}: SHB lineshapes at different magnetic fields, where the
arrow indicates the fixed pump position (1622 meV) and the insets
show the zoomed view of central sharp peaks.}\label{fig_voigt-data}
\end{figure}

The SHB profile changes dramatically in this Voigt geometry. An
ultra-sharp central spike appears on top of the trion spectral
hole, as highlighted by the square region in
Fig.~\ref{fig_voigt-level}(b). The physical origin can be
understood from analysis of the perturbation pathway:
\begin{equation}
\rho_{\bar{x},\bar{x}}^{(0)} \overset{E_{1}^{*}}{\longrightarrow}
\rho_{\bar{x},\bar{t}}^{(1)} \overset{E_{2}}{\longrightarrow}
\rho_{\bar{t},\bar{t}}^{(2)} (
\overset{\Gamma_{t}}{\longrightarrow} \rho_{\bar{x},\bar{x}}^{(2)}
)
 \overset{E_{1}}{\longrightarrow}
\rho_{\bar{t},\bar{x}}^{(3)} \end{equation} The pump field
$E_1^{\ast}$ and probe field $E_2$ create a second order spin
population $\rho^{(2)}_{\bar{x},\bar{x}}$ through the population
decay of the trion, which oscillates at the detuning frequency
$\Delta$ (known as population pulsations \cite{Duncan_PRL1985}).
In the presence of the spin population relaxation process, the
standard OBE predicts a hole burning at $\Delta=0$ with a
linewidth given by the spin relaxation rate $\sim \Gamma_s$
\cite{Duncan_PRL1985}. However, as shown in
Figs.~\ref{fig_voigt-level} and \ref{fig_voigt-data}, the measured
linewidth is orders of magnitude larger than $\Gamma_s$ ($\sim
0.1$ neV) measured by the phase modulation technique for these
QDs~\cite{Cheng_SSC2006}. We will establish below that the extra
broadening is likely to be due to SD processes involving the spin
states.

\begin{figure}[tb] \centering
\includegraphics[width=3.4in]{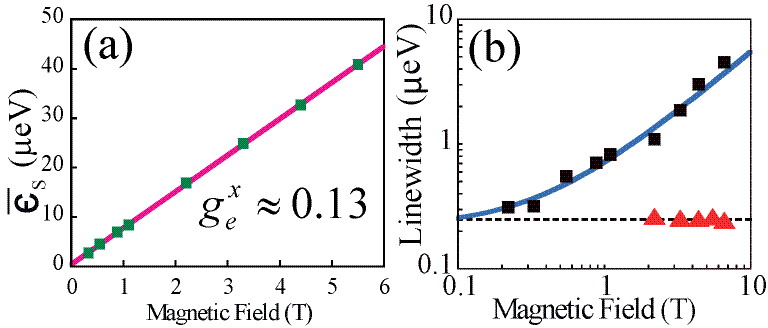} \caption{(a). The
separation between Stokes and anti-Stokes sidebands is plotted as
a function of magnetic field, and the linear fitting gives the
ensemble averaged electron in-plane g-factor $|\bar{g}_e^x|=0.13$.
(b). The linewidth extracted from the experimental data at various
magnetic fields, where $\blacksquare$ ($\blacktriangle$) are the
SHB linewidth of the sidebands (central sharp spike), highlighted
in Fig.~\ref{fig_voigt-level}(b) with ellipse (square) region. The
solid curve is the calculated sidebands linewidth based on
Eq.~(\ref{Eq:SHB-sd3l-spincoh}) (see text). The dashed line is a
guide to the eye.}\label{fig_voigt-Bdep}
\end{figure}

Similar to the trion SD given in Eq.~(\ref{eq:2-level}), the
effects of spin SD processes on the spin population and coherence
are given by,
\begin{eqnarray}
\frac{\partial
\rho_{\bar{x},\bar{x}}(\epsilon_t,\epsilon_s)}{\partial t}
    \bigg | _{\text{SD}} &=&
    \int W_s(\epsilon_t,\epsilon_s;\epsilon_t^{\prime},\epsilon_s^{\prime})
    \rho_{\bar{x},\bar{x}}(\epsilon_t^{\prime},\epsilon_s^{\prime})d\epsilon_t^{\prime} d\epsilon_s^{\prime}  \nonumber \\
    && -\Gamma_s^{SD} (\epsilon_t,\epsilon_s) \rho_{\bar{x},\bar{x}}(\epsilon_t,\epsilon_s) \label{eq:3-level} \\
\frac{\partial \rho_{x,\bar{x}}(\epsilon_t,\epsilon_s)}{\partial t}
    \bigg | _{\text{SD}} &=&
    -\Gamma_s^{SD} (\epsilon_t,\epsilon_s) \rho_{x,\bar{x}}(\epsilon_t,\epsilon_s) \label{eq:3-level_2}
\end{eqnarray}
where the density matrix for each QD in the ensemble is now
characterized by two variables, i.e., the zero field trion
resonance energy $\epsilon_t$ and the spin Zeeman splitting
$\epsilon_s$ in the external magnetic field plus the local field
(e.g., nuclear Overhauser field). Similar to the description of
trion SD, $W_s$ is the redistribution kernel and
$\Gamma_s^{SD}(\epsilon_t,\epsilon_s)=\int
W_s(\epsilon_t^{\prime},\epsilon_s^{\prime};\epsilon_t,\epsilon_s)
d\epsilon_t^{\prime} d\epsilon_s^{\prime}$ is the spin SD rate.
The qualitative feature of the redistribution kernel function
depends critically on the SD mechanism. Local nuclear field
fluctuation induced SD only affects the spin Zeeman splitting:
$W_s =f(\epsilon_s,\epsilon_s^{\prime})
\delta(\epsilon_t-\epsilon_t^{\prime})$. We note that the
inhomogeneous broadening of $\epsilon_s$ induced by the nuclear
field is given by $\Lambda_s \sim 0.1~\mu$eV$\ll \gamma_t$
\cite{Gammon_t2star}. We also note that two quantum dots are
equally excited if the difference in their resonance frequency for
spin to trion transition is much smaller than the trion broadening
$\gamma_t$. Thus, $\rho_{\bar{x},\bar{x}}(\epsilon_t,\epsilon_s) =
\rho_{\bar{x},\bar{x}}(\epsilon_t^{\prime},\epsilon_s^{\prime})$
if $(\epsilon_t-\epsilon_t^{\prime}) \pm
(\epsilon_s-\epsilon_s^{\prime})/2 \ll \gamma_t$. It can then be
shown that the two terms on RHS of Eq.~(\ref{eq:3-level}) cancel
each other and hence this mechanism has a negligible effect on the
linewidth of the ultra-narrow central spike associated with the
spin population pulsation dynamics. On the other hand, if the SD
process is due to the interdot transfer of non-equilibrium spin
population, it is more reasonable to assume a redistribution
kernel $W_s = \Gamma_s^{SD}
\exp[-(\epsilon_s-\bar{\epsilon}_s)^2/\Lambda_s^2-(\epsilon_t-\bar{\epsilon}_t)^2/\Lambda_t^2]/(\pi
\Lambda_t \Lambda_s)$ where $\bar{\epsilon}_t$
($\bar{\epsilon}_s$) is the ensemble averaged trion (spin)
resonance energy \cite{SD_kernel}. In the vicinity of zero
detuning $\Delta \ll \gamma_t$, the DT signal is determined by
\begin{equation}
 E_{\textrm{NL}} \simeq
    \frac{N\sqrt{\pi}|\mu|^4|E_1|^2 E_2^*}{8\hbar^3\Lambda_t(2\gamma_t+\Gamma_t^{SD})}
    \frac{2\Gamma_s+\Gamma_s^{SD}}{\Delta^2+(2\Gamma_s+\Gamma_s^{SD})^2}.
    \label{Eq:SHB-sd3l-spinflip}
\end{equation}
shown as the sharp central spike with linewidth of
$2\Gamma_s+\Gamma_s^{SD}$, in Fig.~\ref{fig_voigt-data}~({\it
Theory}). The experimental data at various magnetic fields are
shown in Fig.~\ref{fig_voigt-data}~({\it Experiment}), and the
measured linewidth of the sharp central spike is plotted in
Fig.~\ref{fig_voigt-Bdep}(b) from which we extract
$\hbar\Gamma^{SD}_s \sim 0.2~\mu$eV.

In addition to the sharp central spike, the newly emerged SHB
features in Voigt geometry also include two symmetric sidebands as
highlighted by the ellipse regions in
Fig.~\ref{fig_voigt-level}(b). These sidebands can be understood
from the perturbation pathway: \begin{equation}
\rho_{\bar{x},\bar{x}}^{(0)} \overset{E_{1}^{*}}{\longrightarrow}
\rho_{\bar{x},\bar{t}}^{(1)} \overset{E_{2}}{\longrightarrow}
\rho_{\bar{x},x}^{(2)} \overset{E_{1}}{\longrightarrow}
\rho_{\bar{t},x}^{(3)} \end{equation} associated with stimulated
Raman spin excitations \cite{Dutt2005}. The Stokes and anti-Stokes
sidebands appear respectively at $\Delta = \pm \mu_B g^x_e B/
\hbar$ and their separation as a function of the magnetic field
gives the ensemble averaged electron g-factor $\bar{g}^x_e \approx
0.13$ (see Fig.~\ref{fig_voigt-Bdep}(a)), in agreement with
earlier reports \cite{Dutt2005}. As the sideband feature is
associated with the Raman spin coherence, it can be inferred from
Eq.~(\ref{eq:3-level_2}) that spin SD process will broaden the
homogeneous linewidth of the sidebands from $\gamma_s$ to
$\gamma_s+\Gamma^{SD}_s$, where $\gamma_s$ is the spin decoherence
rate. The ensemble averaged sideband lineshapes are a convolution
of the homogeneous lineshape with the inhomogeneous broadening of
the spin Zeeman energy. In the vicinity of $\Delta= \pm
\mu_B\bar{g}^x_e B/\hbar$, the nonlinear DT signal is
\begin{eqnarray}
E_{\textrm{NL}} &\simeq&
    \frac{N|\mu|^4|E_1|^2 E_2^* (\gamma_s+\Gamma_s^{SD})}{8 \hbar^3\pi\Lambda_t \Lambda_s \gamma_t(\gamma_t+\Gamma_t^{SD})}
    \nonumber \\
&&    \times \int
    \frac{\text{exp}(-(\epsilon_s - \mu_B\bar{g}^x_e
B)^2/\Lambda_s^2)}{(\gamma_s+\Gamma_s^{SD})^2+(|\Delta|-\epsilon_s)^2}
    d\epsilon_s.
    \label{Eq:SHB-sd3l-spincoh}
\end{eqnarray}
At finite magnetic field, the spin inhomogeneous broadening
$\Lambda_s(B) = \Lambda_s(0) + \mu_B \Delta g^x_e B $ has the
contribution from the inhomogeneity of the electron g-factor
$\Delta g^x_e$ in addition to the nuclear induced zero field
inhomogeneous broadening $\Lambda_s(0)$. The sideband linewidth
measured at various magnetic fields is shown in
Fig.~\ref{fig_voigt-Bdep}(b), which is consistent with that
calculated using Eq.~(\ref{Eq:SHB-sd3l-spincoh}) with parameters
as $\hbar(\gamma_s + \Gamma^{SD}_s) \simeq 0.18~\mu$eV and
$\Lambda_s \simeq 0.64 B~\mu\rm{eV}/ \rm{T} + 0.1~\mu\rm{eV}$. As
$\gamma_s\ll\Gamma_s^{SD}$ from the theoretical investigation of
both the nuclear and phonon induced spin decoherence
~\cite{Yao_PRB2006,Loss_SpinT2_phonon}, we can get
$\hbar\Gamma_s^{SD}=0.18\mu eV$ which agrees with the value of
$\Gamma_s^{SD}$ extracted from the central narrow spike. In
addition, the spin g-factor variation $\Delta g_x^e$ is determined
to be 0.01 from $\Lambda_s$, so $\Delta g_x^e/\bar{g}_x^e \simeq
0.1$, in agreement with transient spin quantum beat
measurements~\cite{Dutt2005}.

In summary, we have shown in this paper that SD from the trion
state complicates the trion SHB profile, and makes an important
contribution to the double-Lorentzian-like lineshape. In the Voigt
geometry, we have found a complex lineshape arising from spin
dynamics in the spectral hole burning: a narrow central spike and
two symmetric Stokes and anti-Stokes sidebands. They have been
theoretically identified to result as consequence of spin
population pulsation dynamics and stimulated Raman spin coherence,
respectively. Moreover, a spin SD process is observed in
contributions to the SHB lineshape, and has been theoretically
identified as interdot transfer of the non-equilibrium spin
population. Possible mechanisms include the interdot spin
flip-flop interactions of the electrons or spin conserved electron
tunneling to adjacent neutral dots. Their quantitive estimates can
not be determined without a detailed calculation which is beyond
the scope of the paper. Nevertheless, the existence of these
decoherence mechanisms revealed by our experiments will have
important impacts on the efforts towards spin based quantum
applications in these kinds of QDs.

\begin{acknowledgments}
This work was supported in part by the U.S. ARO, NSA/LPS, ARDA,
AFOSR, ONR, and FOCUS-NSF.
\end{acknowledgments}

\bibliographystyle{apsrev}

\begin{thebibliography}{28}
\bibitem[{SD_()}]{SD_kernel}
\bibinfo{note}{We have assumed the strong redistribution model for interdot
  trion and spin spectral diffusion where the kernel only depends on the final
  state density~\cite{Berman_1985}.}
\end{thebibliography}

\end{document}